# Solving the multiplication problem of a large language model system using a graph-based method

Turker Tuncer[1], Sengul Dogan[1*], Mehmet Baygin[2], Prabal Datta Barua[3], Abdul Hafeez-Baig[4], Ru-San Tan[5,6], Subrata Chakraborty[7,8], U. Rajendra Acharya[9]

[*1]Department of Digital Forensics Engineering, Technology Faculty, Firat University, Elazig, Turkey
turkertuncer@firat.edu.tr; *sdogan@firat.edu.tr
[2]Department of Computer Engineering, Faculty of Engineering and Architecture, Erzurum Technical University, Erzurum, Turkey
mehmet.baygin@erzurum.edu.tr
[3]School of Business (Information System), University of Southern Queensland, Australia
Prabal.Barua@usq.edu.au
[4]School of Management and Enterprise, University of Southern Queensland, Toowoomba, QLD, Australia
Abdul.Hafeez-Baig@usq.edu.au
[5]Department of Cardiology, National Heart Centre Singapore, Singapore 169609, Singapore
[6]Duke-NUS Medical School, Singapore 169857, Singapore; tan.ru.san@singhealth.com.sg
[7]School of Science and Technology, Faculty of Science, Agriculture, Business and Law, University of New England, Armidale, NSW, 2351, Australia
[8]Center for Advanced Modelling and Geospatial Information Systems, Faculty of Engineering and IT, University of Technology Sydney, Sydney, NSW, 2007, Australia
subrata.chakraborty@une.edu.au
[9]School of Mathematics, Physics and Computing, University of Southern Queensland, Springfield, Australia
Rajendra.Acharya@usq.edu.au

**Abstract:** The generative pre-trained transformer (GPT)-based chatbot software ChatGPT possesses excellent natural language processing capabilities but is inadequate for solving arithmetic problems, especially multiplication. Its GPT structure uses a computational graph for multiplication, which has limited accuracy beyond simple multiplication operations. We developed a graph-based multiplication algorithm that emulated human-like numerical operations by incorporating a ×10k operator, where k represents the maximum power to base 10 of the larger of two input numbers. Our proposed algorithm attained 100% accuracy for 1,000,000 large number multiplication tasks, effectively solving the multiplication challenge of GPT-based and other large language models. Our work highlights the importance of blending simple human insights into the design of artificial intelligence algorithms.

## 1. Introduction

Natural language processing (NLP) enables computer systems to comprehend and generate human languages [1,2]. Large language model (LLM), a form of deep learning trained extensively with text data using NLP methodologies [3], exhibits high performance in language comprehension, generation, and interaction. With advances in artificial intelligence, NLP-based applications for effective and natural communication have proliferated [4,5], e.g., generative pre-trained transformer (GPT), bidirectional encoder representations from transformer (BERT), word2vec, global vectors for word representation (GloVe), robustly optimized BERT pretraining approach (RoBERTa), and text-to-text transfer transformer (T5) [6-8]. BERT analyzes text bidirectionally for deeper understanding of contexts within the language [9]. Based on transformer architecture, GPT is trained on large text datasets and can perform fine-tuning for specific tasks. For instance, GPT-3 has been trained with millions of parameters, which enables it to execute even complex tasks successfully [10]. These models underscore the depth and breadth of how technology has permeated into language understanding and human-machine interactions [11]. Indeed, LLM-based systems have been implemented broadly, from web search engines and virtual assistants to text classification and language translation, offering innovative and ever more efficient and intelligent approaches to information access, content creation, and digital communication [12].

No matter to what degree of complexity an artificial intelligence model has evolved into, the ability to perform basic arithmetic operations remains a core fundamental [13-15]. That said, arithmetic operations that may appear simple to the human mind can be challenging when translated into algorithmic operations [16,17]. Even as chatbots like ChatGPT, which is based on GPT, have revolutionized NLP, their facility with mathematical computation, particularly multiplication, is reportedly inadequate [18,19]. In this article, we examined the multiplication algorithm used by GPT, its strengths and limitations. From analysis of the differences between the GPT and manual approaches to multiplication, we designed a graph-based multiplication algorithm to emulate human-like numerical operations. Improved model performance with the novel algorithm vs. GPT was demonstrated. By studying the precise algorithmic computations underpinning the mathematical intuition driving artificial intelligence-enabled human cognition, we aimed to establish a platform for further research into the subject.

## 2. Proposed graph-based multiplication model

GPT uses a computational graph for multiplication, which has been shown by Dziri et al. [20] to have limited accuracy beyond simple multiplication operations (Figure 1). To solve this problem, we developed a new graph-based algorithm that mimicked human multiplication strategies using a multilevel multiplication operator (Figure 2).

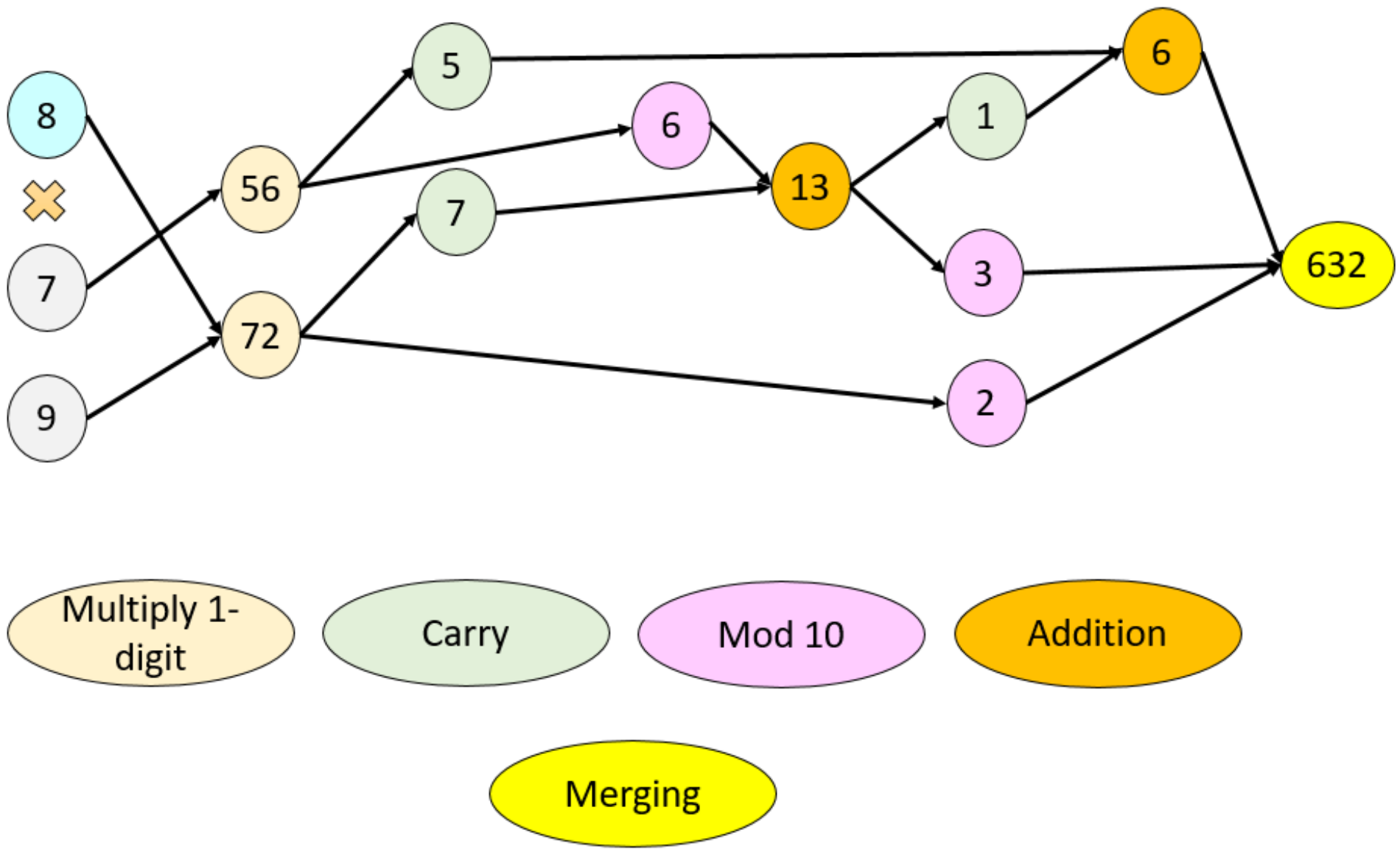

Figure 1. GPT multiplication algorithm. To calculate 8×79 in this example, the algorithm first performs single-digit multiplications (8×7, 8×9), followed by sequential modulo 10, carry, and addition operations. The final correct result, 632, is obtained by merging all penultimate outputs. Model accuracy becomes degraded for multiplication of large numbers [20].

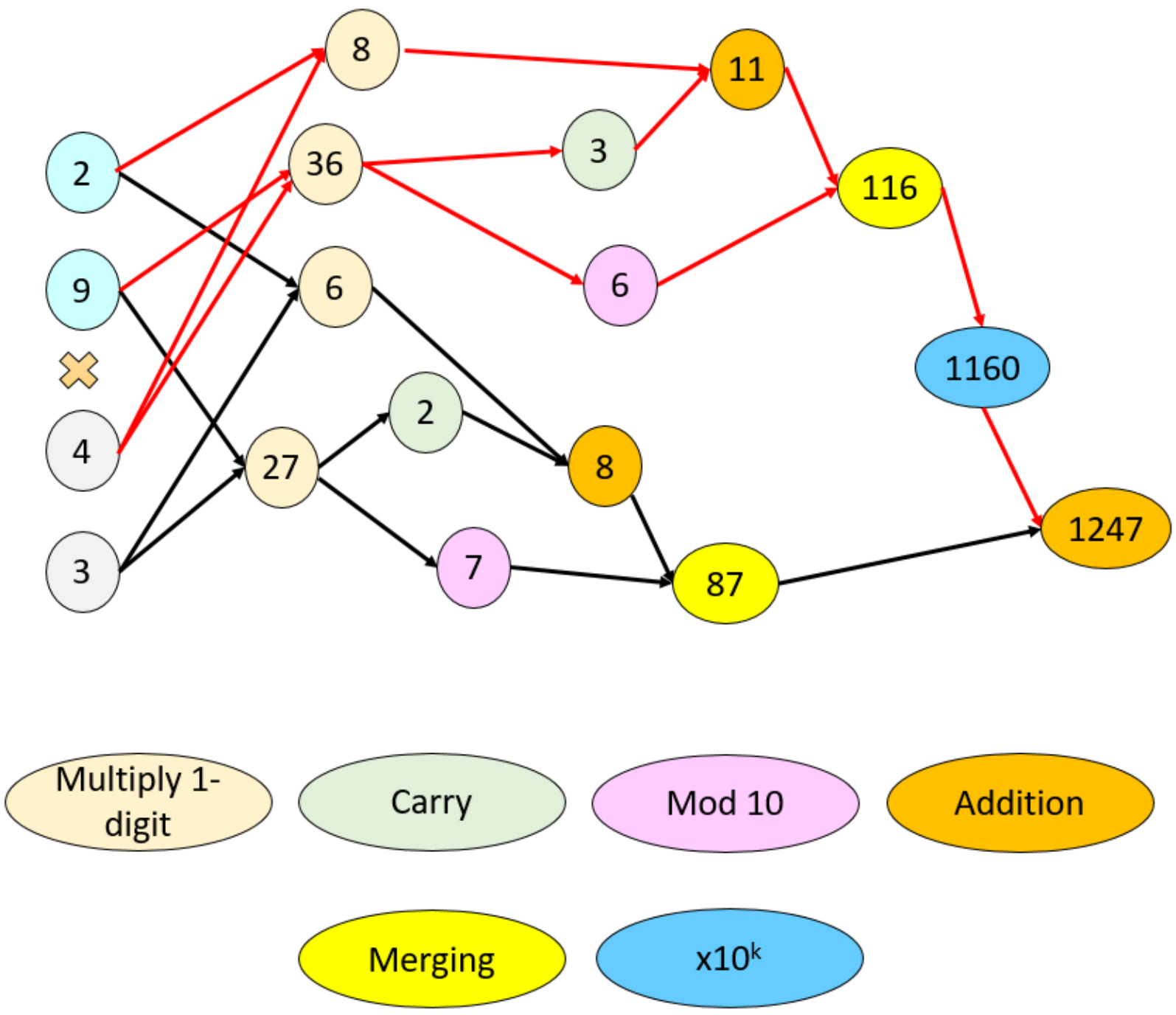

Figure 2. Proposed graph-based multiplication algorithm with added ×10k operator, where k represents the maximum power to base 10 of the larger of two input numbers. To calculate 29×43 in this example (where k=1), the algorithm first performs single-digit multiplication at two levels (0th: 3×9, 3×2; and kth: 4×9, 4×2), followed by sequential modulo 10, carry, and addition operations plus a ×10k operation to the kth level output. The final correct result, 1247, is obtained by merging all penultimate outputs. Black and red arrows indicate processes of the 0th level and kth (k=1) levels, respectively.

The Matlab and Python codes of the proposed model are detailed in Supplementary Tables A1 and A2, respectively. The pseudocode of our proposed solution is given below.

**Step 1:** Initialize two lists representing the numbers to be multiplied.

```
a = [5, 9, 4, 1, 0, 5]
b = [7, 9, 6, 8, 7, 9]
```

The above arrays are sample arrays.

**Step 2:** Create an empty matrix 'tutac' with dimensions (len(a), len(b) + 1) to store intermediate multiplication results and carry values.

```
tutac = np.zeros((len(a), len(b) + 1))
```

**Step 3:** Run multiplication algorithm using nested loops for each digit of 'a' and 'b'.

```
for i in range(len(a) - 1, -1, -1):
    carry = 0
    for j in range(len(b) - 1, -1, -1):
        tt = dizi[a[i], b[j]] + carry // Herein, dizi is the multiplication table.
        number = tt % 10
        carry = tt // 10
        tutac[i, j + 1] = number
    tutac[i, 0] = carry
```

**Step 4:** Calculate individual integer values from the 'tutac' matrix.

```
is_result = np.zeros(len(a), dtype=int)
for i in range(len(a)):
    is_result[i] = sum([int(tutac[i, t]) * 10**(len(tutac[i]) - t - 1)
for t in range(1, len(tutac[i]))])
        is_result[i] += int(tutac[i, 0]) * 10**(len(tutac[i]) - 2)
```

**Step 5:** Combine individual integer values to get the final result.

```
for i in range(len(is_result))])
sonuc = sum([is_result[i] * 10**(len(is_result) - i - 1)
```

Moreover, we have given the pseudocode of the proposed algorithm to better explanation.

Algorithm 1. Pseudocode of the proposed graph-based multiplication algorithm.

```
Input: Input array ‘a’ and ‘b’ and these arrays contain numbers. ‘dizi’ is the multiplaction table.
Output: Multiplication result (result).
---
Initialize matrix ‘tutac’ with dimensions length(a) × (length(b) + 1)
for i from length(a) down to 1 do
    elde = 0
    for j from length(b) down to 1 do
        tt = dizi[a[i] + 1][b[j] + 1] + elde
        sayi = tt modulo 10
        elde = floor(tt / 10)
        tutac[i][j+1] = sayi
    end for j
    tutac[i][1] = elde
    is[i] = 0
    for t from 2 to length(tutac[i]) do
        is[i] = is[i] + tutac[i][t] * 10^(length(tutac[i]) - t)
```

```
    end for t
    is[i] = is[i] + tutac[i][1] * 10^(length(tutac[i]) - 1)
end for i
result = 0
for i from 1 to length(is) do
    result = result + is[i] * 10^(length(is) - i)
end for i
```

## 3. Experimental results

We tested our model on 200,000 problems each of 3-digit × 3-digit, 4-digit × 4-digit, and 5-digit × 5-digit, 3-digit × 4-digit, 3-digit × 5-digit, 4-digit × 5-digit multiplication, i.e., 1,200,000 tasks (for each case, we have used 200,000 tasks) in total. Our graph-based algorithm attained 100% accuracy results for all tasks, compared with accuracies of 59%, 4%, and 0% for 3-digit × 3-digit, 4-digit × 4-digit, and 5-digit × 5-digit multiplication, respectively, as reported for GPT-4 by Dziri et al. [20]. While, our proposed algorithm attained 100% classification performances for the used all 1,200,000 tasks. Table 1 illustrates the discrepant results yielded by GPT-4 vs. our model for example multiplication tasks. As we have added only multiplication with 10k (Tables A1 and A2) to the GPT model, time complexity of our model is equal to $O(m \times n)$ using Big O notation. Where $m$ and $n$ defines the number of the digits of the used inputs.

Table 1. Example multiplication problems and results for proposed model vs. GPT-4. Digit errors in GPT-4 results are highlighted.

| Multiplication | Example problem | Model output (correct) | GPT-4 output |
|---|---|---|---|
| 3-digit × 3-digit | 689×997 | 686933 | 687213 |
| 4-digit × 4-digit | 9247×9019 | 83398693 | 83395093 |
| 5-digit × 5-digit | 10231×48199 | 493123969 | 493301869 |
| 3-digit × 4-digit | 987×8765 | 8651055 | 8653755 |
| 3-digit × 5-digit | 761×98414 | 74893054 | 75000254 |
| 4-digit × 5-digit | 3812×18520 | 70598240 | 70649440 |

Table 1 clearly highlighted that our proposed algorithm enhances the capabilities of existing multiplication algorithms, rendering it a versatile and accurate tool for conducting multiplications on pairs of numbers with unequal lengths.

## 4. Discussion

GPT-4, despite its strength in NLP, has limitations in performing multiplication operations, especially with larger numbers. Its accuracy degraded significantly with the complexity of the multiplication problem, e.g., 0% accuracy for 5-digit multiplication problems. In contrast, our proposed model yielded 100% accuracy results consistently, across 1,200,000 3-, 4- and 5-digit multiplication tasks. By incorporating a ×10k operator into the graph-based multiplication algorithm, our proposed solution emulated human multiplication strategies for base 10 numbers, which rendered the model more relatable, effectively providing an elegant solution to GPT's multiplication problem. To solidify our observations, we compared the performance of our model vs. other LLMs for an example 8-digit × 8-digit multiplication task (Table 2). Our model was the only one that gave the correct answer. The output generated by Bard (Figure 3) —a popular Google LLM-based chatbot application that unlike ChatGPT, which had been trained on a time-limited albeit large dataset, continually draws from data sources in the internet—fared the worst in the comparison (Table 2).

Table 2. Comparison of our model vs. other large language models for the multiplication problem, 99410597×89687949 (8-digit × 8-digit). Digit errors in results are highlighted.

| Model | Model output | Correct |
|---|---|---|
| GPT 3.5 | 8924255330672453 | No |
| GPT 4 | 8915266105078453 | No |
| Bard | 7550948864068302 | No |
| Our model | 8915932553795553 | Yes |

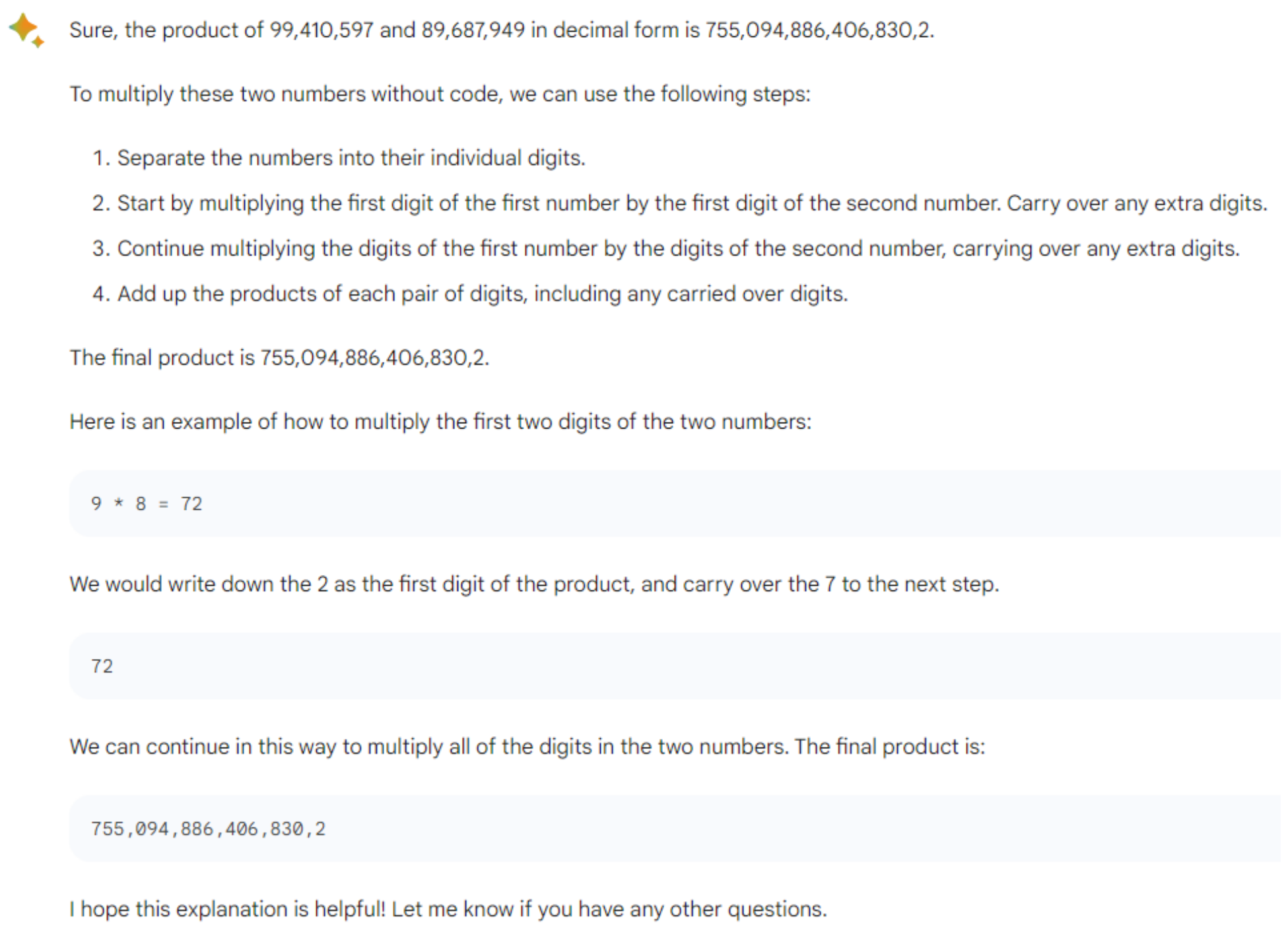

Figure 3. Bard's output for the 99410597×89687949 problem.

Our work primarily focused on the multiplication problem of GPT. The developed model is thus limited to addressing this specific issue and is not generalizable to other arithmetic or computational challenges. In future works, it will be worthwhile to see if the graph-based approach can be extended to other arithmetic operations like division, exponentiation, and root extraction. Our current model exacts a time burden of O(n2). Further optimization can possibly make the model even more efficient, which is pertinent for operations involving very large numbers.

As future work, a seamless integration of this graph-based multiplication model into the ChatGPT framework will be a logical next step. This potentially creates a comprehensive model excelling in both NLP and arithmetic computation. Apart from multiplication, identifying and solving other deficiencies in ChatGPT and like models will be necessary for the development of holistic artificial intelligence models for effective human-computer interaction.

**5. Conclusions**

Despite GPT's advances in NLP, its ability to perform arithmetic calculations, especially multiplication, is suboptimal. This work highlights this gap and has proposed a novel graph-based algorithm based on human multiplication strategies. Our solution attained perfect multiplication results, outperforming state-of-the-art LLM models. The inclusion of a ×10k operator in the multiplication algorithm was instrumental, highlighting the importance of blending simple human insights into complex artificial intelligence algorithms.

**Funding:** This research received no external funding
**Data Availability Statement:** Not applicable.
**Conflicts of Interest:** The authors declare no conflict of interest.

Appendix A

Table A1. Matlab code of the proposed model.

```
clc,clear all,close all
for i=0:9
    for j=0:9
        dizi(i+1,j+1)=i*j;
    end
end
a=[1 0 2 3 1];
b=[4 8 1 9 9];
tutac=zeros(length(a),length(b)+1);
for i=length(a):-1:1
    elde=0;
    for j=length(b):-1:1
        tt=dizi(a(i)+1,b(j)+1)+elde;
        sayi=mod(tt,10);
        elde=floor(tt/10);
        tutac(i,j+1)=sayi;
    end
    tutac(i,1)=elde;
    is(i)=0;
    for t=2:length(tutac(i,:))
        is(i)=is(i)+tutac(i,t)*10^(length(tutac(i,:))-t);
    end
    is(i)=is(i)+tutac(i,1)*10^(length(tutac(i,:))-1);

end
sonuc=0;
for i=1:length(is)
    sonuc=sonuc+is(i)*10^(length(is)-i);
end
result=sonuc;
result
```

Appendix B

Table A2. Python code of the proposed model.

```
import numpy as np

# Initialize lookup table (dizi)
```

```
dizi = np.zeros((10, 10), dtype=int)
for i in range(10):
    for j in range(10):
        dizi[i, j] = i * j

a = [1, 0, 2, 3, 1]
b = [4, 8, 1, 9, 9]
tutac = np.zeros((len(a), len(b) + 1), dtype=int)
is_array = np.zeros(len(a), dtype=int)

# Compute the individual products of digits
for i in range(len(a)-1, -1, -1):
    elde = 0
    for j in range(len(b)-1, -1, -1):
        tt = dizi[a[i], b[j]] + elde
        sayi = tt % 10
        elde = tt // 10
        tutac[i, j+1] = sayi
    tutac[i, 0] = elde

    # Convert the array representation into a number
    for t in range(1, len(tutac[i])):
        is_array[i] += tutac[i, t] * 10**(len(tutac[i]) - t - 1)
    is_array[i] += tutac[i, 0] * 10**(len(tutac[i]) - 1)

# Sum up the results after multiplying by powers of 10 to get the final result
result = 0
for i in range(len(is_array)):
    result += is_array[i] * 10**(len(is_array) - i - 1)
print(result)
```